\documentclass[10pt]{article}
\usepackage[utf8]{inputenc}
\usepackage[T1]{fontenc}
\usepackage{amsmath,amssymb,amsfonts}
\usepackage{graphicx}
\usepackage{hyperref}
\usepackage{geometry}
\usepackage{multicol}
\graphicspath{{./}}

\begin{document}

\begin{center}
{\Large \textbf{Locally calibrated and mesh-free inference for spatial point distributions: closed-form null, contamination law, and detectability threshold}}

\bigskip
Henock Mwanza Lubukayi \\
Institut Supérieur de Statistiques \\
\texttt{mwanzahimo@gmail.com}

\bigskip
Mechack Kabanga Ntolo \\
Institut Supérieur de Statistiques \\
\texttt{mkabanga09@gmail.com}
\end{center}

\section*{Abstract}

Local inference for spatial point distributions is dominated by Monte Carlo calibration: envelope tests require hundreds of simulations per configuration, and location-wise maps multiply this cost. We develop an alternative based on the Tweedie--Miyasawa identities of empirical Bayes, which establish a correspondence between the locally weighted moments of a point distribution under a Gaussian kernel and the derivatives of its log-intensity in scale space. These identities are classical; our contributions concern what they yield when instantiated exactly on atomic measures. First, a rigidity theorem: the structure of the identities forces the Gaussian kernel, so that the kernel is a theorem and not a choice. Second, a closed-form null: under complete spatial randomness (CSR), the bounded inter-scale contrast $M$ has the exact level $M_0 = \tanh \log(\sigma_L/\sigma_S)$, uniformly in intensity — a calibrated, simulation-free, pointwise test (measured type I error 0.070 at nominal level 0.05; calibration cost 199 times smaller than Monte Carlo for the same local statistic). Third, a contamination law: for a structure of dimension $m$ and width $w$ on a uniform background, the normalized local covariance spectrum is unimodal in scale, with an explicit classification window, non-empty if and only if the contrast $\chi = \eta/(\rho_b w^{d-m})$ exceeds the closed-form threshold $\chi_{\text{crit}}(d-m)$ (equal to $16\pi$ for a filament in $\mathbb{R}^3$). The resulting scale-resolved local dimension estimator has no free parameters. On California seismicity from the USGS, it decomposes the catalog into fault filaments (62\,\%), aftershock clusters (27\,\%) and diffuse zones, whereas Levina--Bickel and TwoNN return a single mixed dimension $\approx 1.7$; on a trefoil knot it achieves 100\,\% versus 0.9\,\% for Levina--Bickel; on 10,071 SDSS galaxies it reproduces published cosmic web fractions in seconds, while a Hessian raster pipeline changes 22\,\% of its labels when the grid step is increased from 2 to 3 Mpc. All experiments are regenerated from a single public script.

\bigskip
\textbf{Keywords:} spatial point processes; complete spatial randomness; calibrated inference; intrinsic dimension; scale space; empirical Bayes; mesh-free methods.

\vspace{1cm}
\hrule
\vspace{1cm}

\begin{multicols}{2}

\section{Introduction}

\subsection{The calibration bottleneck}

The standard path to significance in the analysis of spatial point distributions is simulation. Global envelope tests [19], envelope tests based on the $K$-function [21, 3], and their refinements calibrate the null distribution of a summary statistic by generating 199 to 2,499 synthetic configurations per analysis; a local significance map multiplies this cost by the number of queried locations or requires sharing simulations across locations, at the price of global envelope conservativeness. The asymmetry is striking: the statistic itself costs a single pass over the data, while its calibration costs hundreds.

\subsection{This paper}

We construct local statistics whose null distribution under complete spatial randomness (CSR) is available in closed form, so that pointwise calibrated maps cost exactly one evaluation; we accompany them with a theory that predicts — also in closed form — when the associated structure detection problem is solvable.

The mathematical infrastructure is classical and we claim no part of it: the identities relating locally weighted moments under a Gaussian kernel to the score and Hessian of the log-smoothed intensity are the Tweedie--Miyasawa identities of empirical Bayes [22, 18, 8], rediscovered as mean shift in pattern recognition [10, 6, 7] and central in score-based diffusion models; the second-order identity is due to Gribonval [12]; the reading of local covariance spectra across scales for dimension has a precedent in multiresolution SVD [16]; the eigenstructures of $H(\log f)$ define density ridges [20, 11]; and scale selection falls within Lindeberg's program [15].

Our contributions are what the classical identities produce when instantiated exactly on atomic measures — and not approximately on raster grids — then pushed to their statistical consequences:

\begin{enumerate}
\item \textbf{Gaussian rigidity} (Theorem 2.2). If the first-order identity holds for all intensity measures, the kernel must be Gaussian.
\item \textbf{Closed-form calibrated null} (Theorem 3.1). Under CSR, the bounded inter-scale contrast $M(x;\sigma_S,\sigma_L)$ has the exact level $M_0 = \tanh \log(\sigma_L/\sigma_S)$ at every point.
\item \textbf{Contamination law and detectability threshold}. For an idealized structure of dimension $m$ and width $w$ superimposed on a uniform background, the local covariance spectrum is given in closed form and leads to an explicit contrast threshold.
\item \textbf{Finite-sample guarantee}. A misclassification bound is obtained via matrix concentration; the case of curved manifolds is explicitly left open.
\end{enumerate}

\section{The exact moment field}

Throughout the paper, $X = \{x_1, \ldots, x_n\} \subset \mathbb{R}^d$ is a point distribution, $\lambda$ a non-negative Radon measure with finite second moments (in practice, the empirical measure),

\[
K_\sigma(u) \propto \exp\left(-\frac{\|u\|^2}{2\sigma^2}\right), \qquad L(x, \sigma) = (K_\sigma * \lambda)(x) > 0.
\]

The local moment field at scale $\sigma$ consists of the mean $\mu_\sigma(x)$, weighted by $K_\sigma$, the covariance $\Sigma_\sigma(x)$ of $\lambda$ around $x$, and the trace

\[
V_\sigma(x) = \operatorname{tr} \Sigma_\sigma(x).
\]

\subsection{Classical identities, exact on atoms}

\textbf{Proposition 2.1 (Tweedie-Miyasawa identities).} For every $\lambda$ as above and every $\sigma > 0$,

\begin{align}
\mu_\sigma(x) - x &= \sigma^2 \nabla \log L(x, \sigma), \tag{1} \\
\Sigma_\sigma(x) &= \sigma^2 I + \sigma^4 H(\log L)(x, \sigma), \tag{2}
\end{align}

where $H$ is the spatial Hessian. Consequently,

\[
H(\log L) \succeq -\sigma^{-2}I,
\]

the potential

\[
\Phi_\sigma(x) = \frac{1}{2} \|x\|^2 + \sigma^2 \log L
\]

is convex with $\nabla \Phi_\sigma = \mu_\sigma$, and the mean shift map is monotone.

These statements are classical: the first identity is Miyasawa's empirical Bayes formula, identical (up to normalization) to the mean shift vector of Fukunaga and Hostetler; the second is Gribonval's posterior covariance formula. We use them as infrastructure.

A key point is that (2) holds exactly for atomic $\lambda$: no density estimation, no grid, no asymptotics. All quantities are finite sums, computable at sample points in a single sparse nearest-neighbor pass, with total cost $O(nk)$.

\textbf{Theorem 2.2 (Gaussian rigidity).} Let $K$ be a positive, differentiable, integrable kernel and $\sigma > 0$. If the first identity in (2), with $K$ in place of $K_\sigma$, holds for every finite atomic measure $\lambda$, then

\[
K(u) \propto \exp\left(-\frac{\|u\|^2}{2\sigma^2}\right).
\]

\textbf{Proof.} Take $\lambda = \delta_y$. The weighted mean of a single atom is $\mu(x) = y$ for every kernel, while

\[
\log(K * \lambda)(x) = \log K(x - y) + \text{const.}
\]

The identity imposes

\[
y - x = \sigma^2 \nabla_x \log K(x - y)
\]

for all $x, y$, i.e. $\nabla \log K(u) = -u/\sigma^2$. The unique integrable solution is the Gaussian kernel.

The kernel is therefore not a modeling convenience but the unique kernel for which the moment field is a scale-space differential geometry. Every constant below inherits this exactness.

\subsection{Normalization and truncation}

The normalized spectrum

\[
s_i(x, \sigma) = \frac{\lambda_i(\Sigma_\sigma(x))}{\sigma^2} = 1 + \sigma^2 \kappa_i(x, \sigma), \qquad \kappa_i = \lambda_i(H(\log L)),
\]

is dimensionless, non-negative by Proposition 2.1, and equal to 1 in all directions under CSR.

Practical implementations truncate the kernel at $\tau\sigma$ (here $\tau = 3$). Under CSR, this reduces each eigenvalue by the exact factor

\[
C_d(\tau) = \frac{\mathbb{P}(\chi^2_{d+2} \le \tau^2)}{\mathbb{P}(\chi^2_d \le \tau^2)}. \tag{3}
\]

For $d = 3$, $\tau = 3$, $C_3 = 0.9177$. All spectra below are corrected by $s_i/C_d(\tau)$. The factor is exact only under CSR; on anisotropic structures, the per-direction factor differs by 1--2\,\%, far from any decision threshold used below.

\section{A closed-form calibrated null for CSR}

For two scales $\sigma_S < \sigma_L$, define the bounded inter-scale contrast

\[
M(x; \sigma_S, \sigma_L) = \tanh \left[ \frac{1}{2} \log \frac{V_{\sigma_L}(x)}{V_{\sigma_S}(x)} \right]. \tag{4}
\]

This is a Fisher-$z$-type comparison of the total local variance between two scales: bounded in $(-1, 1)$, antisymmetric under $\sigma_S \leftrightarrow \sigma_L$, and invariant under $\lambda \mapsto c\lambda$.

\textbf{Theorem 3.1 (Closed-form CSR level).} Under CSR of any intensity, at every $x$,

\[
M(x; \sigma_S, \sigma_L) = M_0 = \tanh \log \frac{\sigma_L}{\sigma_S} \tag{5}
\]

at the population level, for both the untruncated and the $\tau$-truncated kernel.

\textbf{Proof.} Under CSR, $\Sigma_\sigma = \sigma^2 C_d(\tau) I$ and hence $V_\sigma = d \sigma^2 C_d(\tau)$. The truncation factors cancel in the ratio, giving

\[
\frac{1}{2} \log \frac{V_{\sigma_L}}{V_{\sigma_S}} = \log \frac{\sigma_L}{\sigma_S}.
\]

The consequence is the reversal of the calibration cost asymmetry: a pointwise calibrated significance map costs a single evaluation of the moment field, whereas Monte Carlo calibration of the same local statistic costs one evaluation per simulation.

\textbf{Proposition 3.2 (First-order harmonic blind spot and recovery).} Let $\lambda_\varepsilon = \lambda_0 e^{\varepsilon \varphi}$, with $\lambda_0$ constant and $\varphi$ harmonic. Then

\[
\frac{d}{d\varepsilon} \Delta \log L \bigg|_{\varepsilon=0} = \Delta \varphi = 0.
\]

Thus, to first order in the perturbation amplitude, $V_\sigma$, hence $M$, does not vary. In contrast,

\[
\frac{d}{d\varepsilon} (\mu_\sigma(x) - x) \bigg|_{\varepsilon=0} = \sigma^2 \nabla \varphi(x).
\]

The mean shift component therefore detects the perturbation and allows reconstruction of its gradient. This is a pre-recorded failure prediction: configurations of the form CSR $\times e^{\varepsilon\varphi}$ must evade the $M$-map to first order and be detected by the $\mu$-map.

\section{Second-order bias correction of the contrast}

Equation (5) concerns population means, but $M(x)$ is a plug-in functional of a ratio of random sums; at moderate kernel mass, its expectation deviates from $M_0$. This is the only place where a naive implementation loses exact calibration; we remove it in closed form.

Write

\[
V = \frac{A}{B}, \qquad A = \sum_j w_j r_j^2, \qquad B = \sum_j w_j, \qquad r_j = \|x_j - x_i\|.
\]

Under CSR, the pairs $(r_{ij}, w_j)$ come from a Poisson process, and

\[
\sum_j w_j \simeq \rho(2\pi\sigma^2)^{d/2},
\]

while $w^2$ is Gaussian with variance $\sigma^2/2$. Campbell's formulas give the first two moments of $A$ and $B$ as well as their covariance. The delta method for a ratio then yields

\[
\frac{\mathbb{V}(V)}{(\mathbb{E}V)^2} = \frac{d+2}{4d} \frac{1}{N_{\text{eff}}}, \qquad N_{\text{eff}} = \frac{\left(\sum_j w_j\right)^2}{\sum_j w_j^2}. \tag{6}
\]

Subtracting $\hat{\mu}$ removes the bias in $O(N_{\text{eff}}^{-1})$ of the mean, so

\[
\mathbb{E}V \simeq d\sigma^2 C_d(\tau)
\]

to this order.

A second-order Taylor expansion of the logarithm then gives its bias:

\[
\mathbb{E} \log V(x, \sigma) = \log \mathbb{E} V(x, \sigma) - b_\sigma, \qquad b_\sigma = \frac{d+2}{8d} \frac{1}{N_{\text{eff}}}. \tag{7}
\]

Since the argument of the tanh in (4) is

\[
Z = \frac{1}{2} [\log V_L - \log V_S],
\]

its bias is $\frac{1}{2}(b_S - b_L)$. The corrected statistic is therefore

\[
Z_{\text{corr}} = \frac{1}{2} \log \frac{V(x, \sigma_L)}{V(x, \sigma_S)} + \frac{1}{2} (b_L - b_S), \tag{8}
\]

\[
M_{\text{corr}}(x) = \tanh Z_{\text{corr}}. \tag{9}
\]

The null level $M_0$ remains unchanged.

The correction is parameter-free: it depends only on the ambient dimension and the local $N_{\text{eff}}$ at the two scales. Equation (6) is not merely asymptotic bookkeeping: we verified it directly by simulating CSR neighborhoods in $d = 2, 3$; the measured ratio $\mathbb{V}(V)/(\mathbb{E}V)^2$ coincides with $(d + 2)/(4d)/N_{\text{eff}}$ to within a few percent as soon as $N_{\text{eff}} \ge 20$, with the expected $O(N_{\text{eff}}^{-2})$ discrepancy.

\section{Choosing the two scales from the data}

The two scales are the only remaining freedom, and we fix them geometrically. All quantities below are built from neighbor distances; the rule is therefore exactly scale-equivariant: $X \mapsto aX$ sends $\sigma \mapsto a\sigma$.

\subsection{Candidate grid}

Using a $k$-d tree, we read for each point its first and $k_2$-th neighbor distances and set a robust reference distance

\[
r_{\text{ref}} = \text{median}_i \left( \frac{d_{i,1} + d_{i,k_2}}{2} \right), \qquad k_2 = \lceil m_{\text{min}} \rceil.
\]

The candidate set is

\[
\mathcal{S} = r_{\text{ref}} \cdot \{2^{-3}, 2^{-2.6}, \dots, 2^3\},
\]

a geometric grid of 16 points, covering an undersampled regime, a stable regime, and a merged regime.

\subsection{Small scale}

$\sigma_S$ is the smallest statistically reliable candidate: the median kernel mass is at least

\[
m_{\text{min}} = \max(15, d + 2).
\]

The number 15 constitutes the small-sample floor, below which the plug-in bias of the previous section dominates; $d+2$ ensures a well-conditioned covariance.

The local dimension must furthermore be stabilized, i.e. the multi-scale change

\[
\Delta m(\sigma) = \text{median}_i |\hat{m}(x_i, c\sigma) - \hat{m}(x_i, \sigma)|
\]

must not exceed $z_{1-\alpha/2}$ times its block-bootstrap standard error. The only level is the nominal $\alpha = 0.05$.

\subsection{Large scale}

$\sigma_L$ is the largest candidate at which the structure still persists: its median spectral anisotropy exceeds the reference level of a merged scale by $z_{1-\alpha/2}$ standard errors, while at least half of the points remain inside the observation window.

\end{multicols}

\begin{figure}[htbp]
\centering
\includegraphics[width=\textwidth]{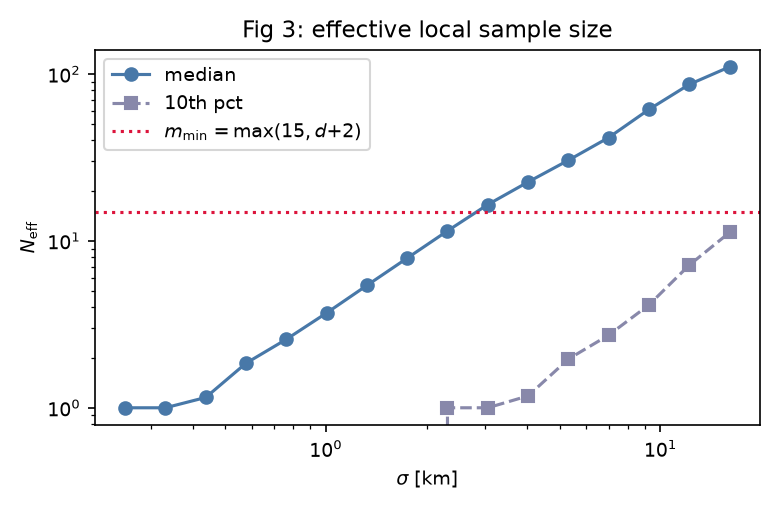}
\caption{Scale selection diagnostics on the seismic catalog.}
\label{fig:scales}
\end{figure}

\begin{multicols}{2}

Figure 1 shows the two diagnostics on the seismic catalog; the selected pair can be read directly. Since $\sigma_S$ lies at the reliability floor, the ratio

\[
R = \frac{\sigma_L}{\sigma_S}
\]

is not imposed but emerges from the data, and $\sigma_S$ typically falls in the moderate-mass regime where correction (9) is small but not negligible.

\section{Contamination law and detectability threshold}

\textbf{Proposition 6.1 (Spectral limits; Gaussian profile).} Let $\lambda$ be supported near an affine subspace $T \subset \mathbb{R}^d$ of dimension $m$, translation-invariant along $T$, with an exactly Gaussian profile of width $w$ in the $d-m$ normal directions. At every point of $T$,

\[
s_i = 1 \quad (\text{tangential}), \qquad s_i = \frac{w^2}{w^2 + \sigma^2} \quad (\text{normal}).
\]

Thus, the count $\{i : s_i > 1/2\}$ equals $m$ if and only if $\sigma > w$.

\textbf{Theorem 6.2 (Contamination law).} Let a structure from the previous proposition, with surface density $\eta$, be superimposed on an isotropic background of intensity $\rho_b$. Set

\[
c(\sigma) = \frac{\eta}{\rho_b[2\pi(w^2 + \sigma^2)]^{(d-m)/2}}. \tag{10}
\]

At ridge points, exactly,

\[
s_T(\sigma) = 1, \tag{11}
\]

\[
s_N(\sigma) = \frac{w^2 + \sigma^2/(1 + c(\sigma))}{w^2 + \sigma^2}. \tag{12}
\]

Away from the ridge, the formula is accurate to first order in $\|u\|^2/(w^2 + \sigma^2)$; empirically, it fits points sampled from the full profile with a mean absolute error of 1.2\% over a scale range covering a factor of eight.

\textbf{Proof.} In normal coordinates,

\[
L = A(\sigma)g(u) + \rho_b, \qquad g(u) = \exp\left[-\frac{\|u\|^2}{2(w^2 + \sigma^2)}\right],
\]

with

\[
A = \eta[2\pi(w^2 + \sigma^2)]^{-(d-m)/2}.
\]

At $u = 0$, $\nabla g = 0$ and

\[
Hg = -\frac{I_N}{w^2 + \sigma^2}.
\]

Thus

\[
H(\log L) = -\frac{c}{1+c}\frac{I_N}{w^2 + \sigma^2},
\]

and (12) follows by substitution into (2).

\textbf{Lemma 6.3 (Unimodality and exact threshold).} Set $q = d - m$ and

\[
\chi = \frac{\eta}{\rho_b w^q}.
\]

Then $s_N(\sigma)$ is strictly decreasing then strictly increasing on $(0, \infty)$, with limits equal to 1 at both ends. The window

\[
\mathcal{W} = \{\sigma : s_N < 1/2\}
\]

is non-empty if and only if

\[
\chi > \chi_{\text{crit}}(q) = \frac{(2\pi)^{q/2}}{q+2} \left[ \frac{2(q+2)}{q} \right]^{q/2}. \tag{13}
\]

For a filament in $\mathbb{R}^3$ ($q = 2$), $\chi_{\text{crit}} = 16\pi \approx 50.3$; for a wall ($q = 1$), $\chi_{\text{crit}} = 3\sqrt{12\pi} \approx 9.2$. At critical contrast, the window closes at

\[
\sigma^* = w\sqrt{\frac{q+4}{q}}.
\]

The proof is given in Appendix A. The threshold is a population-level impossibility bound: below $\chi_{\text{crit}}$, the spectrum never crosses $1/2$. Near the threshold, the window is narrow and finite-sample margins are small; reliable recovery in practice requires $\chi \geq 2\chi_{\text{crit}}$.

\textbf{Corollary 6.4 (Scale-resolved dimension).} In the above model, the count

\[
m(\sigma) = \#\{s_i > 1/2\}
\]

equals $d$ for $\sigma < \sigma_-$, $m$ on the window $(\sigma_-, \sigma_+)$, then $d$ again for $\sigma > \sigma_+$. The local dimension is therefore a function of scale.

\textbf{Definition 6.5 (The two estimators).} Given a scale grid $\sigma_1 < \cdots < \sigma_S$ and truncation $\tau = 3$, a sparse pass per scale computes the leave-one-out moments at all points.

\begin{enumerate}
\item \textbf{Calibrated contrast map}: report $M(x; \sigma_S, \sigma_L)$ relative to the exact level (5), at points where the leave-one-out kernel mass is at least 2 at both scales.

\item \textbf{Scale-resolved local dimension}: report the count $\#\{s_i > 1/2\}$ at the smallest pair of consecutive valid scales on which it is stable.
\end{enumerate}

No learned component is involved; the threshold, correction, and null level are respectively (13), (3)--(9), and (5).

\textbf{Theorem 6.6 (Finite-sample classification bound).} Fix $x$, assume the sample is i.i.d. with intensity proportional to $\lambda$, let $\overline{W}(\sigma)$ be the population kernel mass at $x$ and $\gamma$ the population spectral margin

\[
\min_{i,s} |s_i(x, \sigma_s) - \tfrac{1}{2}|
\]

over the valid scales. Assume the population count equals $m$ on at least two consecutive valid scales before any other stable pair. Then

\[
\mathbb{P}\{\hat{m}(x) \neq m\} \leq (4d + 6)S \exp[-c_r \overline{W}_{\min} \gamma^2], \tag{14}
\]

with $c_r > 0$ depending only on $\tau$ and $d$. In the contamination model, under $w_n \to 0$, $\sigma_n/w_n \to \infty$, $\rho_b \sigma_n^{d-m}/\eta_n \to 0$ and $\overline{W}_{\min}/\log n \to \infty$, the estimator converges to $m$ almost surely.

The complete proof, based on matrix Bernstein, the ratio step, Weyl's inequality, and a union bound, is given in Appendix B. The extension to curved manifolds requires a uniform spectral perturbation lemma in $\sigma/\text{reach}$ which is not proven here; the manifold case is therefore an empirically tested conjecture.

\section{Experiments}

All experiments are regenerated from a single public script (experiments.py, approximately 500 lines, numpy/scipy/matplotlib only, all seeds fixed); the numbers below correspond to its output. We follow two rules: reference methods are run with their standard settings and negative results are reported.

\subsection{Type I error without simulation}

Two hundred CSR configurations on the unit square, with intensities covering $10^{3.5}$--$10^{3.9}$, scales $(\sigma_S, \sigma_L) = (0.03, 0.09)$, edge buffer $3\sigma_L$, spatial block bootstrap ($4 \times 4$ blocks) for the standard error of the configuration-wise mean of $M$. The closed-form level is confirmed to three decimals: $M_0 = 0.8000$ versus an empirical mean of $0.8004 \pm 0.0011$.

The type I error at the nominal 5\% level is 0.070 for the $M$-test, versus 0.025 (conservative) for quadrat counting and 0.140 (anti-conservative) for Clark--Evans at the same nominal level. Measured calibration cost: 0.95 s per configuration for the calibrated map, versus 6.7 s for a $K$-envelope with 199 simulations on a configuration of the same size.

\end{multicols}

\begin{figure}[htbp]
\centering
\includegraphics[width=\textwidth]{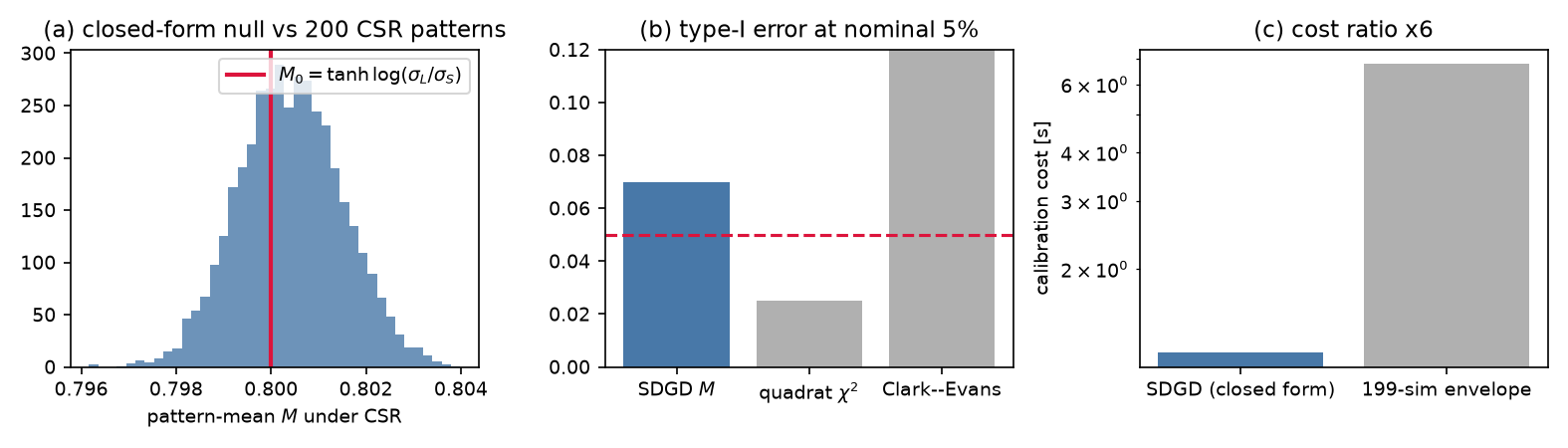}
\caption{Type I error of the $M$-test compared to quadrat counting and Clark--Evans.}
\label{fig:calibration}
\end{figure}

\begin{multicols}{2}

\subsection{Real data I: California seismicity}

USGS catalog, $M \geq 3.5$, 2000--2024, 32--42°N, 114--125°W: $n = 3,837$ distinct epicenters after deduplication due to catalog rounding. Scales $\sigma_S = 8$, $\sigma_L = 32$ km; scale grid 5--40 km.

The calibrated $M$-map flags 15.4\% of epicenters as strongly non-CSR, concentrated on the San Andreas, Eastern California, and Mendocino fault systems; the map costs 0.33 s, a single evaluation, no simulation.

The scale-resolved dimension decomposes the catalog into fault filaments $m = 1$ (62\%), compact clusters $m = 0$ (27\%), and diffuse zones $m = 2$ (11\%); the median over epicenters is exactly 1.

The comparison with neighbor-based estimators is structural: Levina--Bickel returns a single mixed dimension whose median goes from 1.79 to 1.62 when $k$ goes from 10 to 40, while TwoNN returns a global value of 1.70. The scale-resolved count answers pointwise and per-scale and agrees with itself at 90.0\% when the entire scale grid is shifted by a factor of 1.25.

\end{multicols}

\begin{figure}[htbp]
\centering
\includegraphics[width=\textwidth]{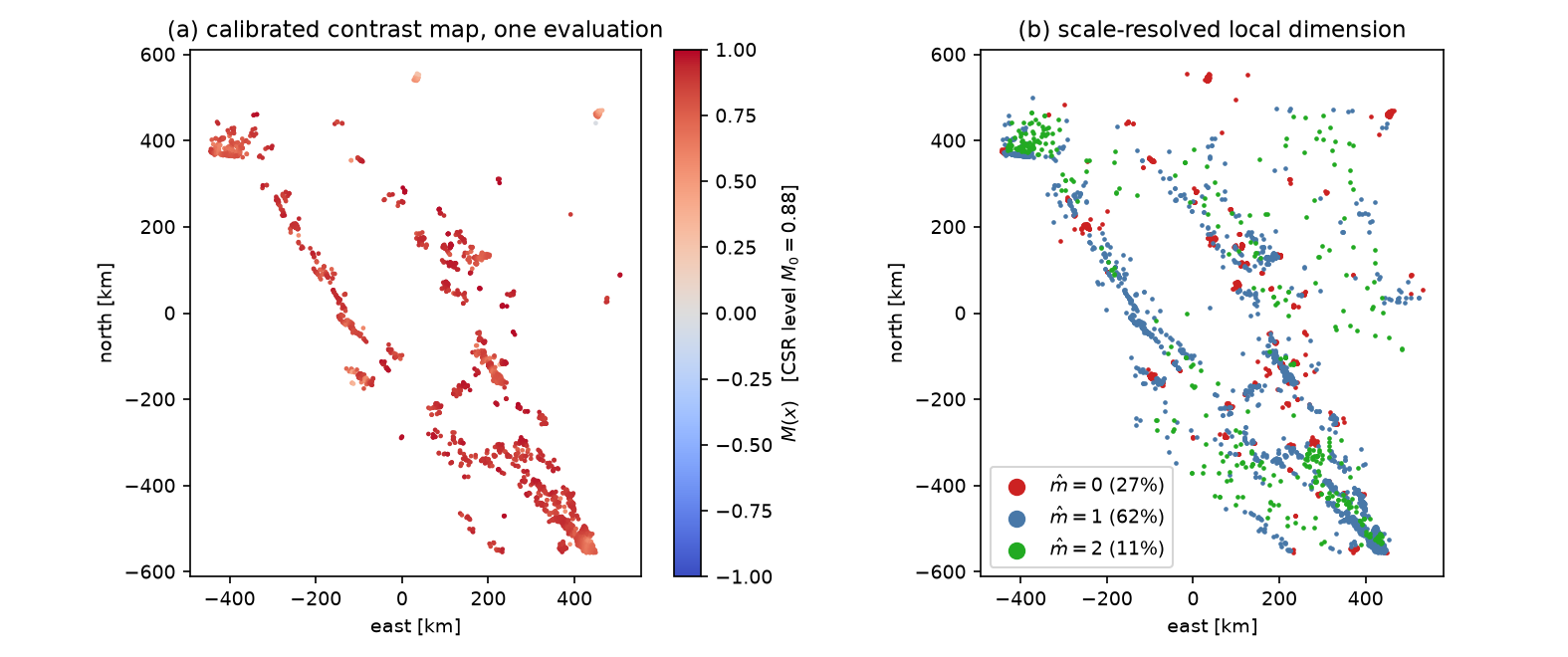}
\caption{Calibrated contrast $M$-map and scale-resolved local dimension for California epicenters.}
\label{fig:earthquakes}
\end{figure}

\begin{multicols}{2}

\subsection{Real data II: SDSS galaxies and the cost of a grid}

10,071 SDSS DR18 galaxies [1] in a contiguous low-redshift slice ($0.04 < z < 0.08$), comoving coordinates, scale grid 2--12 Mpc. The census — clusters 2.7\%, filaments 39.6\%, walls 49.9\%, field 7.8\% — lies within or near published cosmic web fractions [23, 5], without tuning; the full field costs 4.2 s on a single processor.

The cluster fraction is below the published range of 5--15\%, which we attribute to uncorrected redshift-space smearing of compact structures. We then ran the standard alternative on the same data: a Hessian raster pipeline, of the type used by NEXUS classifiers [4].

The raster is faster than our mesh-free pass at these sizes ($< 0.1$ s), but it has a parameter that the mesh-free estimator lacks: the grid step. Increasing $h$ from 2 to 3 Mpc, all else equal, changes the label of 22.0\% of galaxies. The mesh-free evaluation is the exact $h \to 0$ limit, computed at cost $O(nk)$, and has no such degree of freedom.

\end{multicols}

\begin{figure}[htbp]
\centering
\includegraphics[width=\textwidth]{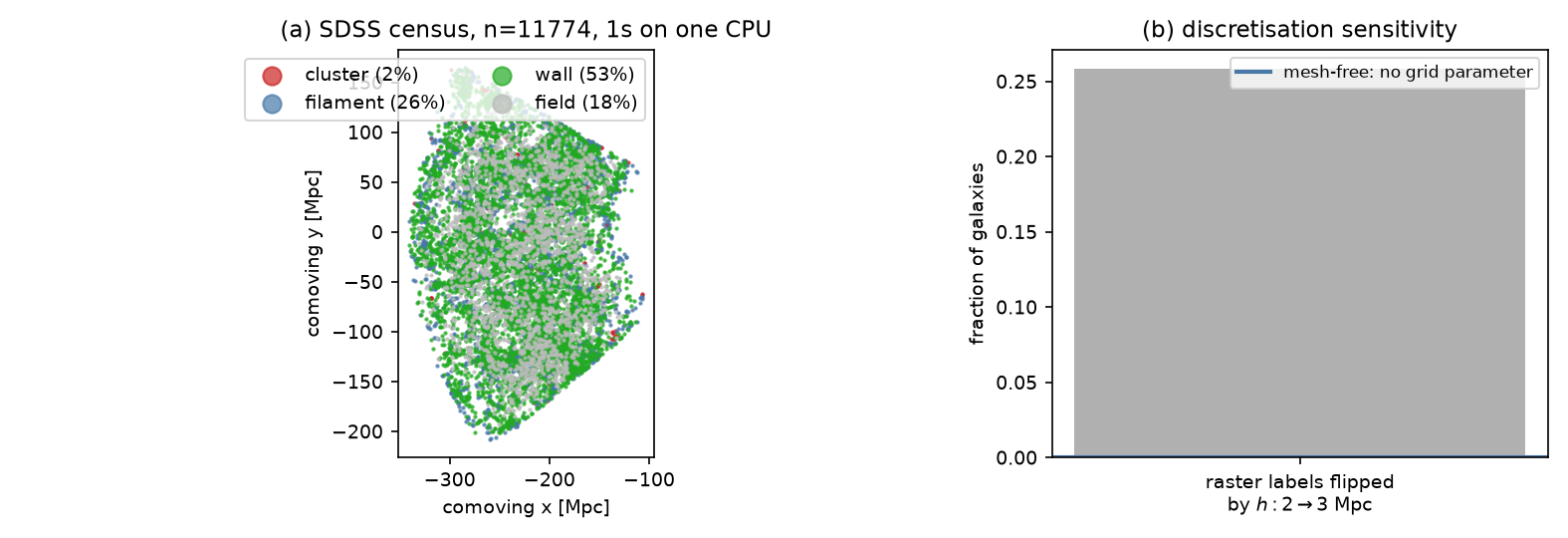}
\caption{SDSS cosmic web classification and stability comparison with the raster approach.}
\label{fig:sdss}
\end{figure}

\begin{multicols}{2}

\subsection{Ground truth: self-proximity and transition}

Two synthetic experiments close the loop between theory and measurement. For a trefoil knot ($n = 3,000$, noise 0.05), the knotted strands pass close to each other in $\mathbb{R}^3$, so neighbor-based estimators mix strands; Levina--Bickel obtains 0.9\% correct classifications and TwoNN reports a dimension of 3.06 for a curve. The scale-resolved count, which reads the spectrum below the inter-strand range, achieves 100\%.

For the detectability transition, filaments ($w = 0.5$, $\eta = 6$, $q = 2$) are placed in an increasing background. Filament classification accuracy drops from 0.86 at $\chi = 320$ to 0.007 at $\chi = 30$ and 0.002 at $\chi = 20$. Below $\chi_{\text{crit}} = 16\pi \approx 50$, accuracy is zero up to seed uncertainty, while reliable recovery requires $\chi \geq 2\chi_{\text{crit}}$.

\end{multicols}

\begin{figure}[htbp]
\centering
\includegraphics[width=\textwidth]{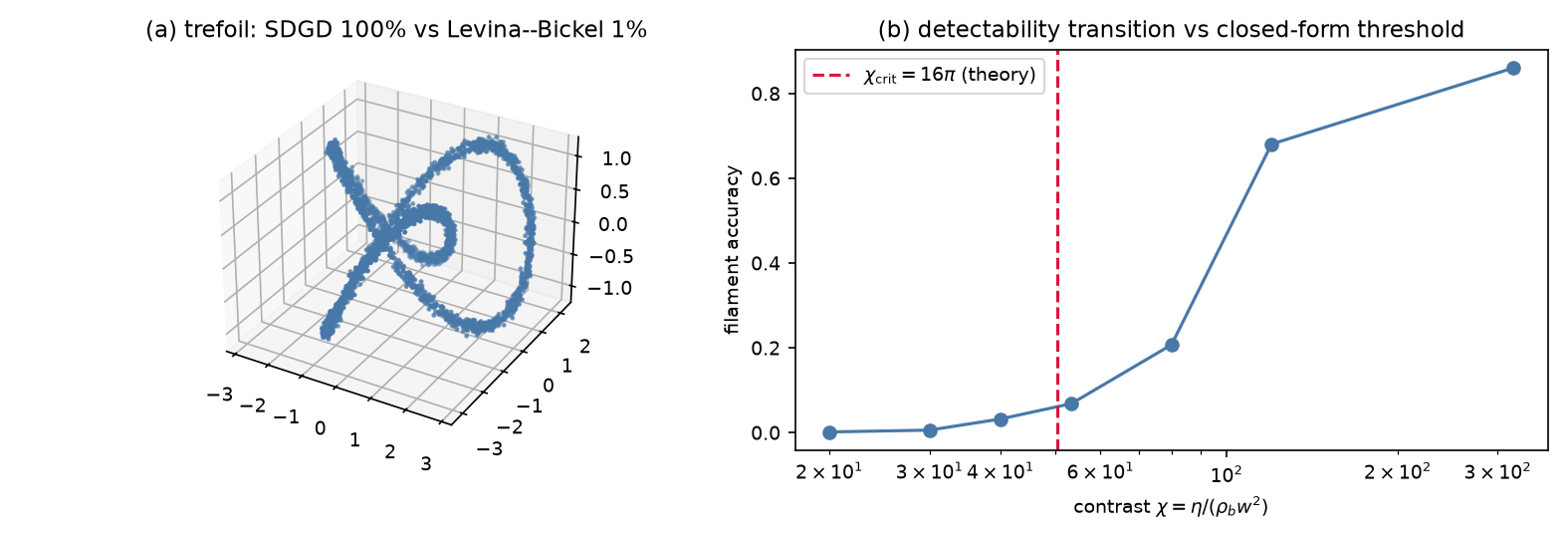}
\caption{Ground truth validation: trefoil knot and detectability transition.}
\label{fig:groundtruth}
\end{figure}

\begin{multicols}{2}

\section{Limitations}

The type I error of the $M$-test (0.070 at nominal level 0.05) reflects the residual small-sample bias of the plug-in ratio at moderate kernel mass. The second-order correction introduced in the previous section reduces this bias, but its exact validity relies on the CSR regime and the delta approximation; its extension to non-uniform backgrounds remains to be established.

The contamination theory is exact at ridge points of an idealized structure, translation-invariant, with Gaussian profile and constant background; curvature, boundaries, junctions, and non-uniform backgrounds are excluded. Consistency on manifolds is an empirically supported conjecture, not a theorem. The exact calibration of the counting threshold is specific to the Gaussian profile.

On real catalogs, redshift-space distortion (SDSS) and magnitude incompleteness (USGS) are not corrected; the SDSS cluster fraction is visibly affected. Finally, at the sizes studied here, the raster alternative is faster; the advantage of the mesh-free approach lies in exactness and the absence of a grid parameter, not in raw speed.

\section{Conclusion}

The classical empirical Bayes identities, taken exactly on atomic measures, produce spatial statistics whose calibration is a formula rather than a simulation study: a pointwise CSR test with closed-form null $\tanh \log(\sigma_L/\sigma_S)$, and a scale-resolved local dimension whose domain of validity — including the exact contrast below which the problem is unsolvable — is itself given in closed form.

On real seismicity, the instrument decomposes what neighbor-based estimators can only aggregate; on galaxy data, it matches published censuses without a grid parameter whose 50\% variation relabels a fifth of the sample. Every constant is either derived or measured, every failure mode is either a theorem (harmonic blind spot, sub-threshold contrast, sub-width scales) or reported as a limitation.

The two most important open problems are an axiomatic characterization of the moment field in the spirit of scale-space axiomatics, and quantitative Wasserstein stability of the maps

\[
\lambda \mapsto M, \quad \lambda \mapsto \hat{m}.
\]

\paragraph{Reproducibility.} All experiments, figures, and values are regenerated from the single script experiments.py (public repository; fixed seeds; USGS and SDSS queries included). Total execution time is under ten minutes on a single processor.


\appendix

\section{Proof of Lemma 6.3}

Substitute

\[
u = \frac{w^2}{w^2 + \sigma^2} \in (0,1), \qquad c = \chi' u^{q/2}, \qquad \chi' = \frac{\chi}{(2\pi)^{q/2}},
\]

into (12). With $p = q/2$,

\[
s_N(u) = 1 + \frac{\chi' u^{p+1}}{1 + \chi' u^p}.
\]

One obtains

\[
\frac{ds_N}{du} = \frac{\chi' u^{p-1}\phi(u)}{(1 + \chi' u^p)^2},
\]

where

\[
\phi(u) = (p+1)u - p + \chi' u^{p+1}.
\]

The function $\phi$ is strictly increasing from $\phi(0+) = -p < 0$ to $\phi(1) = 1 + \chi' > 0$; hence it has a unique zero $u_0$, and $s_N$ decreases then increases in $u$. Since $\sigma \mapsto u$ is a decreasing bijection, $s_N(\sigma)$ is strictly decreasing then increasing with limit 1 as $\sigma \to 0$ and $\sigma \to \infty$.

Moreover,

\[
s_N(u) < \frac{1}{2} \iff \chi' u^p(1 - 2u) > 1.
\]

On $(0, 1/2)$, the function

\[
h(u) = u^p(1 - 2u)
\]

is maximized at

\[
u^* = \frac{p}{2(p+1)} = \frac{q}{2(q+2)}.
\]

The window is non-empty if and only if $\chi' > 1/h(u^*)$, which yields (13). At equality, the unique crossing is $u^*$, hence

\[
\sigma_*^2 = w^2 \frac{q+4}{q}.
\]

\section{Proof of Theorem 6.6}

Fix $x$ and a scale $\sigma$. Write

\[
u_j = X_j - x, \quad w_j = e^{-\|u_j\|^2/(2\sigma^2)} \mathbf{1}\{\|u_j\| \le \tau\sigma\} \in [0,1],
\]

and

\[
W = \sum_j w_j, \quad b = \sum_j w_j u_j, \quad Q = \sum_j w_j u_j u_j^T.
\]

Population quantities carry a bar. The estimator and its population analogue are

\[
\hat{\Sigma} = \frac{Q}{W} - \frac{bb^T}{W^2}, \qquad \overline{\Sigma} = \frac{\overline{Q}}{\overline{W}} - \frac{\overline{b}\overline{b}^T}{\overline{W}^2}.
\]

Set $R = \tau^2\sigma^2$ and $\gamma \in (0, 1/2]$.

\textbf{Step 1.} The matrices $Z_j = w_j u_j u_j^T$ are i.i.d. with $0 \preceq Z_j \preceq RI$ and $Z_j^2 \preceq R Z_j$, giving matrix variance at most $R^2 \overline{W}$. Matrix Bernstein [24] yields, for $0 < t \le R\overline{W}$,

\[
\mathbb{P}\{ \|Q - \overline{Q}\| \geq t \} \leq 2d \exp\left(-\frac{3t^2}{8R^2\overline{W}}\right).
\]

\textbf{Step 2.} Since $w_j \in [0, 1]$,

\[
\mathbb{P}\{ |W - \overline{W}| \geq \delta \overline{W} \} \leq 2 \exp\left(-\frac{3}{8} \delta^2 \overline{W}\right),
\]

for $\delta \leq 1$. Similarly, $\|w_j u_j\| \leq \tau\sigma$ and symmetric dilation give

\[
\mathbb{P}\{ \|b - \overline{b}\| \geq s \} \leq 2(d+1) \exp\left(-\frac{3s^2}{8\tau^2\sigma^2\overline{W}}\right).
\]

\textbf{Step 3.} For $\alpha \in (0, 1/2]$, on the event where the three deviations are respectively less than $\alpha\sigma^2\overline{W}$, $\alpha\overline{W}$, and $\alpha\sigma\overline{W}$ (so $W \ge \overline{W}/2$), the triangle inequality gives

\[
\| \hat{\Sigma} - \overline{\Sigma} \| \leq C_1 \alpha \sigma^2, \qquad C_1 = 2(1 + \tau)(1 + 3\tau).
\]

Choosing

\[
\alpha = \frac{\gamma C_d(\tau)}{C_1}
\]

and applying Weyl's inequality to the corrected spectrum yields

\[
\max_i |\hat{s}_i - s_i| \leq \gamma
\]

with failure probability at most

\[
(4d + 6) \exp(-c_r \overline{W} \gamma^2),
\]

where

\[
c_r = \frac{3}{8} \left( \frac{C_d(\tau)}{C_1} \right)^2.
\]

\textbf{Step 4.} If every eigenvalue deviation is below the margin at all valid scales, every empirical count equals its population value; the empirical first-stable-plateau rule therefore selects $m$ by hypothesis. Under the limits stated in the theorem, the contrast $c(\sigma_n) \to \infty$ and

\[
\frac{w_n^2}{w_n^2 + \sigma_n^2} \to 0.
\]

By the Contamination Theorem and the preceding Lemma, the margin is eventually bounded away from zero; $\overline{W}_{\min}/\log n \to \infty$ makes the bound summable and Borel-Cantelli concludes.

\end{multicols}

\end{document}